%% file: IEEE-conference-template-062824.tex
\documentclass[conference]{IEEEtran}
\IEEEoverridecommandlockouts
\usepackage{cite}
\usepackage{amsmath,amssymb,amsfonts}
\usepackage{algorithmic}
\usepackage{graphicx}
\usepackage{textcomp}
\usepackage{xcolor}
\usepackage{booktabs}
\usepackage{url}
\usepackage{subfigure}
\usepackage{dirtytalk} 
\def\BibTeX{{\rm B\kern-.05em{\sc i\kern-.025em b}\kern-.08em
T\kern-.1667em\lower.7ex\hbox{E}\kern-.125emX}}

\usepackage{pgfplots}
\pgfplotsset{compat=1.18}
\usepgfplotslibrary{groupplots, fillbetween}
\usetikzlibrary{decorations.pathreplacing}
\usetikzlibrary{positioning, calc, shapes.geometric, arrows.meta}

\begin{document}

\title{StePPO: Adaptive Step-Size Control for Initial-Value Problems Using Proximal Policy Optimisation}

\author{\IEEEauthorblockN{1\textsuperscript{st} Milan Koumans}
\IEEEauthorblockA{\textit{Signal Processing Systems} \\
\textit{Eindhoven University of Technology}\\
Eindhoven, The Netherlands \\
\textit{m.koumans@tue.nl}}
\and
\IEEEauthorblockN{2\textsuperscript{rd} Nicola Pezzotti}
\IEEEauthorblockA{\textit{SCAN } \\
\textit{ASML} \\
Eindhoven, The Netherlands \\
\textit{nicola.pezzotti@asml.com}}
\and
\IEEEauthorblockN{3\textsuperscript{nd} Tristan S.W. Stevens}
\IEEEauthorblockA{\textit{Signal Processing Systems} \\
\textit{Eindhoven University of Technology}\\
Eindhoven, The Netherlands \\
\textit{t.s.w.stevens@tue.nl}}}

\maketitle

\begin{abstract}
ODE solvers are used to model, simulate, and analyse continuous-time systems found in many signal processing applications. Time integration of these initial-value problems is traditionally performed using adaptive embedded Runge--Kutta methods. For proper time-step selection, adaptive solvers accept or reject each candidate step according to a local error estimate. Their efficiency is therefore controlled by the rule that selects the next step size. Classical proportional--integral--derivative (PID) controllers are computationally inexpensive per decision and reliable, but their reactive design can lead to costly numerical solves near stiff transitions. We formulate step-size selection for parametrised IVPs as a Markov decision process and train a continuous-action policy using proximal policy optimisation. The policy processes its observation signals using a deep network while maintaining the embedded solver logic. We evaluate on a decay function, the Van der Pol system, and the Brusselator system at matched tolerances, and compare against a standard PID controller and a multi-scale DeepONet network. We find that the proposed learned policy remains within prescribed error tolerances while reducing the required number of solve steps by up to $60\%$, achieving a wall-clock time reduction of $58\%$, for both in- and out-of-distribution settings.
\end{abstract}


\section{Introduction}
Solving initial value problems (IVPs) is central to many fields of science and engineering, including physics \cite{guzman_numerical_2023}. Moreover, it is also an important component for data-driven methods, such as generating high-quality training data \cite{navarro_solving_2024, cao_laplace_2024, lu_deeponet_2021, de_florio_physics-informed_2024}. Given an ordinary differential equation (ODE) with a known initial state and system parameter, one typically solves this using adaptive embedded Runge--Kutta ($RK$) methods, which combine multiple evaluations of the local vector field to determine the update and an embedded local error estimate \cite{butcher1964implicit}. Given a fixed error tolerance, the sequence of step sizes during an integral solve determines whether a solve is cheap or expensive: excessively large steps waste computation on rejected attempts, whereas overly conservative steps require unnecessarily many steps. This trade-off becomes particularly important when the local dynamics vary strongly along the trajectory. Stiff systems provide a prominent example, as their widely separated time scales can cause the appropriate step size to change substantially throughout the integration interval. Efficient integration therefore depends not only on satisfying the local error tolerance, but also on adapting step sizes rapidly to changes in the numerical behaviour of the system.

This adaptation is conventionally performed by proportional--integral--derivative (PID) controllers \cite{soderlind2003digital, soderlind_automatic_nodate}. Classical PID controllers are reactive by construction: their feedback law is fixed in advance and performs well on smooth systems, but they cannot respond to (sufficiently large) changes in numerical behaviour before local error evidence from within that region has already been observed, and a rejected step is only ever corrected \emph{after} the wasted computation has already occurred. Moreover, PID control has a limited observation space, namely the local error (or its recent history), and is therefore constrained to infer information only from this limited, causal observation. When expanding the observation space, possibly capturing state-dependent behaviour, one could inform a controller for a specific (family of) ODEs with a learned model, due to its broader model capacity. Two relevant paradigms are: \emph{surrogate} models that predict a state given system coordinates and \emph{adaptive} controllers that adjust the step online from feedback observed during the solve. Here, we restrict focus to initial-value problems for ODEs.

Surrogate approaches learn a function mapping coordinates (e.g. time) directly to state, most notably in physics-informed neural networks (PINNs) \cite{raissi_physics-informed_2019}. Once trained, inference can be substantially faster compared to classical solvers, and they have been applied widely to both partial differential equations \cite{luo_physics-informed_2025} and ODEs \cite{de_florio_physics-informed_2024}. Their central limitation is generalisation: a PINN is learned for a single instance and does not extend to, or is very limited for, a distribution of decay coefficients without retraining for specific coefficients; it moreover struggles on systems with sharp transients \cite{krishnapriyan_characterizing_nodate}. Neural operators, such as DeepONet, address this by learning a function-to-function map rather than a single solution \cite{lu_deeponet_2021, lu_deepxde_2021, liu2021multiscale}, and have been used to solve differential equations directly \cite{cao_laplace_2024}. However, this broader generality typically requires expressive models and substantial training data, which is itself costly, while accurately solving functions exhibiting sharp transitions remains challenging \cite{lu_deeponet_2021}.

A different approach is to keep the embedded solver with step-size control, and optimise step direction \cite{goodship_dynamic_2025, ouala_learning_2021}, step size, or both \cite{liu_hierarchical_nodate}. We choose to focus on step size due to the widespread availability of preconfigured solvers, and because it keeps the same measured local error, which allows us to directly compare to current algorithms. For this approach, the classical solver is in full control of the update and we learn only its step-size controller; that is, we replace the fixed PID feedback law with a learned one. Learning a controller that generalises across a distribution of task instances requires a model capable of relating the observed numerical state to the consequences of candidate step-size decisions. This relation is difficult to obtain through supervised learning: there is generally no labelled data for a particular solver state, and the quality of a decision depends not only on its immediate outcome but also on how it affects the subsequent integration trajectory. The numerical outcome of a candidate step is not available to the controller before that step is evaluated. We therefore treat the outcome of the solver as an unknown transition process and learn the step-size policy through model-free reinforcement learning (RL), using observed transitions and rewards. A clear reward signal can be defined from observations without requiring ground-truth data, and the resulting policy can be trained across a distribution of task instances rather than fit to a single trajectory.

To this end, we learn a continuous-action policy that adapts the step size from the observed solver state aiming for a single policy that generalises across a distribution of parametrised IVPs. To address learning on a task distribution, we augment the instantaneous observation with a latent representation learned from the preceding interaction history. This representation provides the policy with information about trajectory- and task-dependent behaviour that is not available from the current observation alone. Our main contributions are as follows.
\begin{itemize}
    \item We formulate adaptive IVP step-size control as a sequential decision problem in which a learned continuous-action policy selects the step size while the embedded numerical solver retains full authority over state evolution and local-error acceptance.
    \item We augment the policy with a learned representation of the observed solver trajectory, allowing it to condition step-size decisions on temporal information beyond the instantaneous solver state.
    \item We evaluate the proposed hybrid approach against pure model- and data-driven methods on both non-stiff and stiff ODEs, including task-parameter interpolation and extrapolation, and assess both accuracy and computational cost.
\end{itemize}

\section{Related work}
The use of RL for adaptive time stepping is recent and has seen relatively few contributions. One example selects a step size from a predefined candidate set using deep Q-learning \cite{dellnitz_efficient_2023}. The authors also fit the output weights $b_j$ of an explicit $RK$ scheme by linear regression for a given problem class. Both forms of adaptation are restricted to structured parametrisations: the action space is intrinsically discrete, and because the authors set $\gamma = 0$ the objective is greedy, so the controller cannot trade a locally expensive step against future savings. The explicit integrator further confines admissible step sizes to a narrow stability domain, whereas we target implicit schemes with a correspondingly wider admissible range. Continuous-action RL has also been used for step-size selection in nonsmooth systems \cite{riley_reinforcement_2025}. Their policy directly selects the next time step and incorporates the estimated local error into a reward that trades accuracy against computational cost. Consequently, the learned policy participates in determining the achieved accuracy; more specifically, the reward shapes the speed--accuracy trade-off during training. In contrast, we treat RL strictly as an adaptive controller around an otherwise unchanged embedded IVP solver: proposed steps remain subject to the solver's conventional tolerance-based acceptance criterion. We further formulate this controller over a distribution of parametrised IVPs and investigate whether trajectory-conditioned policies can exploit information beyond the local error signal and generalise to unseen task parameters.

Neural operators are an adjacent paradigm that aims to reduce the computational cost of repeatedly solving differential equations by learning a direct map from inputs to solutions, rather than using a numerical integrator. Given input coordinates together with a task parameter, a neural operator produces output values in a single forward pass; this pass is not autoregressively dependent on the coordinates, which allows the full trajectory to be computed in parallel. Neural operators have been trained on ODEs \cite{cao_lno_2023, li_improving_2025}, demonstrating their ability to approximate parametrised trajectories. Because our focus is on stiff systems, this provides a relevant comparison to adaptive numerical integration. Since \say{vanilla} neural operators can struggle with sharp transitions, we employ a multiscale architecture that aims to exploit oscillatory behaviour to compensate for this \cite{liu2021multiscale}.

\section{Background}
\label{sec:method}

\input{figures/1_figure_method}

\subsection{Initial Value Problem using adaptive time stepping}
\label{sec:background_ivp}
An initial value problem (IVP) consists of an ODE together with an initial condition: given $t_0$ and $\mathbf{y}_0$, find $\mathbf{y}(t)$ such that $\dot{\mathbf{y}}(t)=f\!\left(t,\mathbf{y}(t)\right)$ and $\mathbf{y}(t_0)=\mathbf{y}_0$. In general, $\mathbf{y}(t)$ is not available in closed form, so one resorts to numerical integration. A time integrator discretises the continuous variable $t$ into solver steps $i$ with times $t_i$ and numerical states $\mathbf{y}_i \approx \mathbf{y}(t_i)$. An $RK$ method evaluates the vector field at several intermediate stages within each step and combines these evaluations to obtain an order-$p$ approximation, with local truncation error $\mathcal{O}(\Delta t_i^{p+1})$. For stiff systems, explicit $RK$ methods may require prohibitively small step sizes to remain stable; hence implicit methods are commonly used \cite{butcher1964implicit}. Fig.~\ref{fig:method}(a) illustrates this discretisation of $y(t)$ into numerical states $y_n$.

In the absence of an analytical solution, there is no ground truth to compare against. Instead, a proposed step is accepted if it lies within a prescribed error tolerance. This is typically done by comparing the primary embedded approximation
\begin{equation}
    \mathbf{y}_{i+1}^{(p)} = \mathbf{y}_i + \Delta t_i \cdot \mathbf{d}_i^{(p)}
\end{equation}
to a second embedded approximation
\begin{equation}
    \mathbf{y}_{i+1}^{(\hat{p})} = \mathbf{y}_i + \Delta t_i \cdot \mathbf{d}_i^{(\hat{p})}
\end{equation}
where $p$ and $\hat{p}$ denote the orders of the two methods. Their difference gives $\boldsymbol{\epsilon}_{i}=\mathbf{y}^{(p)}_{i+1}-\mathbf{y}^{(\hat{p})}_{i+1}$. We then define the normalised error of element $k$ for step $i$ as
\begin{equation}
    e_{i}^{(k)} = \frac{\epsilon_{i}^{(k)}}{\text{a}_{\text{tol}} + \text{r}_{\text{tol}} \cdot \max\!\left(|y_{i}^{(k)}|, |y^{(p,k)}_{i+1}|\right)}
\end{equation}
where $y_{i}^{(k)}$ is the current state component, and $\text{a}_{\text{tol}}/\text{r}_{\text{tol}}$ are the absolute and relative tolerances defined for each system. Since we have a multi-dimensional problem, we take the root-mean-square error
\begin{equation}
    e_{i} = \sqrt{\frac{1}{d}\sum_{k=1}^{d} \left(e_{i}^{(k)}\right)^2}
\label{eq:e_i}
\end{equation}
to transform this multi-dimensional vector into a scalar. This provides a scalar normalised embedded error estimate (error ratio) for the local step. Crucially, the step $\mathbf{y}_{i+1}$ is accepted whenever $e_{i} \leq 1$ and rejected otherwise; a rejected step is retried from the same $t_{i}$ with a reduced step size. As shown in Fig.~\ref{fig:method}(b), the proposed step depends directly on $\Delta t$, which requires a complete recalculation of $e_{i}$.

In adaptive time stepping, $\Delta t_{i+1}$ is adapted during an integral solve and given by $\Delta t_{i+1} = f(e_{i})$. Choices for $f(\cdot)$ can be based on heuristics or feedback loops \cite{soderlind_time-step_2006}. The latter is widely adopted, for example via PID controllers. The PID controller uses $e_{i}$ together with the error history to control $\Delta t$. Given the default configuration $k_{p} = 0$, $k_{i} = 1$ and $k_{d} = 0$, the controller reduces to the classical elementary (integral) controller
\begin{equation}
    u_{i} = \operatorname{clip}\!\left( \eta \, e_{i}^{-1/\hat{p}},
        \; u_{\min}, \; u_{\max} \right).
    \label{eq:i-controller}
\end{equation}
where $\hat{p}$ denotes the order of the error estimate. The resulting step-size update is
\begin{equation}
    \Delta t_{i+1}=u_i\Delta t_i.
\end{equation}
Intuitively, a small error ($e_{i} \ll 1$) yields $u_{i} > 1$ and the step grows, while a large error ($e_{i} > 1$) yields $u_{i} < 1$ and the step shrinks; the factor $\frac{1}{\hat{p}}$ ties the rate of adaptation to the convergence order of the error estimate. As shown in Fig.~\ref{fig:method}(a), a PID step is not necessarily optimal; $h^*$ indicates the optimal step that still lies within the error tolerances. As such, there is ample room to improve such a controller.

\subsection{Initial value problems as a Markov Decision Process}
\label{sec:background_ivp_as_mdp}
Numerical integration of an IVP can be interpreted as a sequential state-transition process. Starting from $(t_0,\mathbf{y}_0)$, the solver iteratively advances the numerical state until $t\geq t_{\text{end}}$ or the budget $N_{\text{steps}}$ is exhausted. We refer to such a full sequence of states as a trajectory $\tau$. At each iteration, the outcome of the next solver attempt is determined by the current numerical state and the proposed step size. We denote the complete numerical state of the integration process at step $i$ by $s_i$, including $t_i$, $\mathbf{y}_i$, the previous step size $\Delta t_{i-1}$, and any relevant controller history. For a fixed numerical integration scheme, the subsequent state depends only on the current complete state, the proposed step size, and the task parameter $\xi$. Consequently, the integration process satisfies the Markov property,
\begin{equation}
p(s_{i+1}\mid s_0,\ldots,s_i,a_{i},\xi) = p(s_{i+1}\mid s_i,a_{i},\xi).
\label{eq:markov}
\end{equation}
where $a_{i}$ is the multiplicative step-size action. For IVP parametrisation $\xi$ and numerical scheme $RK$, the solver induces an action-conditioned transition
\begin{equation}
s_{i+1} = \mathcal{T}_{\xi}(s_i,a_i).
\end{equation}
Although this transition is deterministic for the systems considered here, its outcome is not available to the controller at decision time. In particular, the candidate state $\mathbf{y}_{i+1}$, the corresponding local error estimate $e_{i}$, and the eventual accept/reject decision are obtained only after evaluating the proposed numerical step. Evaluating $\mathcal{T}_{\xi}(s_i,a_i)$ is therefore precisely the computation whose cost the step-size controller seeks to use efficiently.

Classical adaptive controllers do not explicitly predict this action-conditioned transition. Instead, they select the next step size from quantities observed during previous solver evaluations, most prominently the local error history. The resulting control law is therefore primarily reactive: whether the newly proposed step was appropriate becomes clear only after that transition has been evaluated. This motivates a controller that can exploit the observed trajectory to anticipate the likely numerical consequence of a candidate step before executing it. The controller observation is a projection of the complete numerical state; in particular, the current step size need not be included in the observation even though it affects the transition.

\subsection{Reinforcement learning}
\label{sec:background_rl}
Reinforcement learning aims to learn a stochastic policy
\begin{equation}
a_i \sim \pi_{\theta}(a_i \mid o_i),
\label{eq:policy}
\end{equation}
where $o_i$ denotes the observation available to the controller and $\theta$ denotes the learnable policy parameters \cite{sutton2018reinforcement}. The environment evolves from state $s_i$ according to $s_{i+1}=\mathcal T(s_i,a_i;\xi)$, where $\xi$ denotes task parameters that may vary between environments. The task parameter $\xi$ may be supplied as part of $o_i$, but the controller need not have access to the complete environment state $s_i$; instead, it selects its action from the available observation $o_i$. Each transition produces a subsequent observation $o_{i+1}$ and reward $r_{i+1}$. Given the sequence of state transitions (trajectory), we can describe a complete sequence as
\begin{equation}
    \tau = (o_0,a_0,r_1,o_1,\ldots,a_{N-1},r_N,o_N),
\end{equation}
with $N$ the sequence budget. The objective is to maximise the expected return over the complete trajectory,
\begin{equation}
    J(\theta)=\mathbb{E}_{\tau\sim\pi_{\theta}}\left[\sum_{i=0}^{N-1}\gamma^i r_{i+1}\right]
\label{eq:J}
\end{equation}
with discount factor $\gamma\in[0,1]$.

Directly attributing (\ref{eq:J}) to individual actions is non-trivial; as such, we optimise the policy using Proximal Policy Optimisation (PPO) \cite{schulman_proximal_2017}. PPO follows an actor--critic formulation, where the actor $\pi_\theta(a_i\mid o_i)$ defines the action distribution and the critic $V_\phi(o_i)$ estimates the expected future return used as a baseline for evaluating sampled actions.

This difference is represented through the advantage estimate. We employ generalised advantage estimation (GAE) ~\cite{schulman_high-dimensional_2018}
\begin{align}
\delta_i &= r_{i+1} + \gamma V_\phi(o_{i+1}) - V_\phi(o_i), \\
\hat{A}_i &= \sum_{l=0}^{N-i-1} (\gamma\cdot\lambda_{\mathrm{GAE}})^l\cdot\delta_{i+l},
\label{eq:gae}
\end{align}
where $\lambda_{\mathrm{GAE}}\in[0,1]$ controls the bias--variance trade-off of the estimate. The advantage estimate determines how the probability density of the sampled action should be adjusted: actions with $\hat A_i>0$ performed better than expected and should become more likely, whereas actions with $\hat A_i<0$ should become less likely. PPO incorporates this signal through the clipped surrogate objective
\begin{equation}
L^{\mathrm{clip}}(\theta) = \mathbb{E}_i \left[ \min\left( \rho_i(\theta)\hat A_i,\, \operatorname{clip} \left( \rho_i(\theta),1-\epsilon,1+\epsilon \right)\hat A_i \right) \right],
\end{equation}
where
\begin{equation}
\rho_i(\theta) = \frac{\pi_\theta(a_i\mid o_i)}{\pi_{\theta_{\mathrm{old}}}(a_i\mid o_i)}
\end{equation}
is the probability-density ratio for the sampled action.

The critic is simultaneously fitted to the return target
\begin{equation}
\hat R_i = \hat A_i + V_\phi(o_i),
\end{equation}
where $\hat A_i$ and $\hat R_i$ are computed from the collected rollout and treated as fixed targets during the subsequent PPO update. This target is used in
\begin{equation}
L^V(\phi) = \frac{1}{N} \sum_i \left[ V_\phi(o_i)-\hat R_i \right]^2.
\end{equation}
Finally, an entropy term $L^{\mathrm{ent}} = \mathbb{E}_i[H[\pi_\theta(\cdot\mid o_i)]]$ encourages exploration during training. We maximise the complete objective
\begin{equation}
L = L^{\mathrm{clip}} -c_1L^V +c_2L^{\mathrm{ent}},
\end{equation}
where $c_1$ and $c_2$ respectively weight the value and entropy terms.

\input{figures/2_trajectories}

\section{Methods}
\label{sec:methods}
Our central design is based on separating numerical accuracy from computational efficiency. Specifically, the embedded solver remains responsible for the state update and for enforcing numerical accuracy, while our controller merely focuses on efficient solver step proposals. Consequently, the policy cannot improve efficiency by relaxing the
error criterion. Classical controllers determine the next step proposal from error estimates. We instead hypothesise that the numerical consequence of a candidate step can be better anticipated by exploiting a richer description of the current trajectory and task. We therefore learn a controller that maps the available solver information to a multiplicative
step-size factor $g_i$
\begin{equation}
    \Delta t_{i}=g_i\Delta t_{i-1},
\end{equation}
with $a_i=g_i$ and as such $\pi_\theta:o_i\mapsto a_i=g_i$. Given the expected complexity of this mapping, we use a deep neural network $f_{\theta}(\cdot)$ with learnable parameters $\theta$. To learn such a policy, we must define a reward function. A desirable policy reaches $t_{\text{end}}$ using as few solver evaluations as possible while respecting the prescribed numerical tolerance. We therefore define the relative progress through the integration interval as
\begin{equation}
\Phi(t_i) = \min\!\left(\frac{t_i-t_0}{t_{\mathrm{end}}-t_0},1\right),
\end{equation}
such that
\begin{equation}
\Delta\Phi_i = \Phi(t_{i+1})-\Phi(t_i)
\end{equation}
measures the progress resulting from a solver attempt. An accepted step yields positive progress, whereas a rejected step leaves $t$ unchanged and consequently gives $\Delta\Phi_i=0$. We additionally reward completion with unused computational budget,
\begin{equation}
    r_{i+1} = \Delta \Phi_i + \mathbf{1}_{\{t_{i+1} = t_{\mathrm{end}}\}} \cdot c_{\mathrm{term}}\left(1 - \frac{N_{i+1}}{N_{\max}}\right)
    \label{eq:reward}
\end{equation}
where $N_i$ is the number of solver attempts used (including rejected attempts), and $N_{\max}$ denotes the trajectory budget. Trajectories terminate upon reaching $t_{\mathrm{end}}$ or exhausting the budget; no separate failure penalty is applied, and the terminal coordinate is constrained such that $t_i\leq t_{\mathrm{end}}$. Notably, the local error $e_i$ does not enter the reward explicitly: the local-error acceptance criterion remains enforced by the embedded error estimator of the numerical solver.

Interaction with the solver follows the conventional reset/step formulation. We define an episode as one complete integration from $(t_0,\mathbf{y}_0)$ until either $t_{\mathrm{end}}$ is reached or the step budget is exhausted. At the beginning of an episode, a task parameter $\xi$ is sampled and the numerical state is initialised at $(t_0,\mathbf{y}_0)$. The observation supplied to the policy contains only quantities available during numerical integration,
\begin{equation}
    o_i  = \left[ \tilde{\mathbf{y}}_{i}; \tilde{e}_{i-1}; \widetilde{\Delta e}_{i-1}; \alpha_{i-1}; \xi \right].
\end{equation}
Here, $\tilde{\mathbf{y}}_i$ is the normalised solution state, $\Delta e_{i-1} = e_{i-1} - e_{i-2}$ is the error difference, $\tilde e_{i-1}$ and $\widetilde{\Delta e}_{i-1}$ denote the corresponding normalised error features, and $\alpha_{i-1}\in\{0,1\}$ indicates whether the previous solver attempt was accepted.

Tracking trajectory dynamics is expected to have a positive effect on performance. We therefore use an encoder--decoder alongside the policy to maintain a latent state $z_i$ summarising the interaction history. The policy and value function are conditioned on this embedding, $\pi_\theta(a_i\mid o_i,z_i)$ and $V_\phi(o_i,z_i)$. The embedding provides trajectory context that is useful for control; it is not interpreted as an inferred task parameter. The task parameter $\xi$ is supplied to the policy and used as an auxiliary decoder target to provide a task-specific learning signal to the latent space. We train the variational auto-encoder on a buffer \(\mathcal{D}=\{\tau_i\}\), similar to VariBAD \cite{zintgraf2020varibad}. We use a four-dimensional latent state with encoder widths \([128,128]\) and decoder widths \([128,64,32]\). The decoder objectives \(r_{\text{reward}}\), \(r_{\text{accept}}\), \(r_{\text{task}}\), and \(r_{\text{state}}\) are weighted \(0.45\), \(0.30\), \(0.20\), and \(0.05\), respectively, where \(r_{\text{task}}\) and \(r_{\text{state}}\) are our additions.

\input{figures/3_figure}

\section{Experiment}

\subsection{ODE systems}
We evaluate three ODE systems with varying task parameters: scalar decay, the Van der Pol oscillator \cite{van_der_pol_lxxxviii_1926}, and the Brusselator \cite{prigogine_symmetry_1968}. The scalar decay problem is
\begin{equation}
    \dot{y}=-\mu y,
\end{equation}
where increasing $\mu$ imposes increasingly restrictive stability requirements. The Van der Pol oscillator is given by
\begin{equation}
    \begin{aligned}
    \dot{y}_1 &= y_2,\\
    \dot{y}_2 &= \mu(1-y_1^2)y_2-y_1,
    \end{aligned}
\label{eq:vdp-system}
\end{equation}
with state $\mathbf{y}=(y_1,y_2)^\top$ and nonlinearity parameter $\mu$. Its state-dependent dynamics and sharp transitions can make PID control inefficient. The Brusselator is
\begin{equation}
    \begin{aligned}
    \dot{y}_1 &= A+y_1^2y_2-(B+1)y_1,\\
    \dot{y}_2 &= By_1-y_1^2y_2,
    \end{aligned}
\label{eq:brusselator}
\end{equation}
where $A=1$ and $B\in[1,50]$. Its stiffness depends on the parameter regime and trajectory. We denote the generic task parameter by $\xi$, corresponding to $\mu$ for the first two systems and to $B$ for the Brusselator.

To solve the trajectories, we use \texttt{Kvaerno5} from \texttt{Diffrax}, with the \texttt{Diffrax} PID controller as the baseline adaptive controller \cite{kidger_neural_2022}. We use the elementary controller $k_{p} = 0$, $k_{i} = 1$ and $k_{d} = 0$ with $\eta=0.9$ and $[u_{\min},u_{\max}]=[0.2,10]$. Moreover, we include an \say{oracle} controller to provide an approximate efficiency reference. Starting from a PID step, it probes larger accepted steps using relative gains of $25\%$, $12.5\%$, $6.25\%$, and so on. If the PID step is rejected, the candidate step is reduced by $25\%$ until recovery or termination. Rejected candidates do not replace the best accepted state, so the oracle is an approximate reference rather than a computational lower bound.

\subsection{PPO}
All agents are trained for \(750\) PPO iterations across \(N_{\text{env}}=256\) parallel environments, using rollout horizons \(T=50\), \(750\), and \(1200\) for scalar decay, Van der Pol, and Brusselator, respectively. We use \(\gamma=0.99\), \(\lambda_{\mathrm{GAE}}=0.95\), a clip range of \(0.2\), value coefficient \(c_1=0.5\), and entropy coefficient \(\beta_H=0.005\) \cite{andrychowicz_what_2020}. Each iteration uses four PPO epochs with four minibatches per epoch. The policy uses hidden widths \([128,64,32]\) with \(\tanh\) activations. Adam is initialised at \(10^{-3}\) and annealed to \(10^{-5}\). For the latent variant, the VAE is updated twice per iteration from the same collected trajectories using the same learning-rate schedule. We run four seeds per configuration.

\subsection{DeepONet}
We adopt a custom DeepONet implementation based on the Cartesian-product formulation \cite{lu_deeponet_2021} as implemented in DeepXDE \cite{lu_deepxde_2021}, extended with a multi-scale Fourier feature trunk following the multiscale DeepONet formulation for oscillatory operator learning described for seismic wave response modelling \cite{liu2021multiscale}. Input coordinates are projected through two parallel random Gaussian Fourier features at scales $\sigma= {1,10}$ before entering shared trunk layers. Both branch and trunk networks use five hidden layers of width 256 with ReLU activations and Glorot normal initialisation, trained jointly with Adam for $1$k epochs on $8192$ training and $8192$ held-out test trajectories with batch size $256$. We have observed a plateau during training well before the number of epochs was reached.

We simulate the systems over $[0,10]$, $[0,50]$, and $[0,100]$, respectively, using $r_{\text{tol}}=10^{-3}$ and $a_{\text{tol}}=10^{-6}$. Fig.~\ref{fig:trajectories} shows the trajectories and local error estimates $e_i$ from (\ref{eq:e_i}) of PID (elementary), StePPO and DeepONet; proposed steps with $e_i>1$ are rejected for PID and StePPO and therefore omitted from the trajectories. We consider the integrated error
\begin{equation}
    e_{\text{int}} = \left( \frac{1}{t_{\text{end}} - t_0} \right) \sum_{i=1}^{N-1} \frac{\hat e(i) + \hat e(i-1)}{2}
\end{equation}
where $\hat e(i)$ is the relative error compared to the reference trajectory.


\section{Results}
\label{sec:results}

Figure~\ref{fig:results} shows the integrated error $e_{\text{int}}$ and number of steps $n_{\text{steps}}$ for the decay, Van der Pol, and Brusselator systems. DeepONet produces substantially larger errors than the adaptive controllers under the prescribed error tolerances and varying task dynamics. PID and StePPO achieve comparable errors, with small differences arising from deviations of the numerical trajectories relative to $y_{\text{ref}}$.

For the decay problem, StePPO initially requires more steps than PID. Given the short integration interval $t_{\text{end}}=10.0$ and low decay rates $\xi<5$, the potential reduction in solver steps is small. Although some runs outperform PID across the task distribution, this improvement is not consistent. For both Van der Pol and Brusselator, StePPO substantially reduces the number of steps required and approaches the local oracle reference. This reduction is achieved through fewer rejected steps and larger accepted step sizes. The learned controller therefore adapts more effectively in regions where the admissible step size changes rapidly, reducing the number of costly rejected attempts. Nevertheless, StePPO does not achieve oracle controller performance, indicating theoretical room for improvement.


\subsection{Robustness of learned models.}
PID is task-agnostic by construction, whereas learned models may exhibit reduced performance outside their training support. We therefore consider two training protocols on the Van der Pol system: a \emph{full-range} setting, trained on $\xi \in [1,50]$, and a \emph{binned} setting, trained only on $\xi \in [5,15] \cup [35,45]$. All models are evaluated on $\xi \in [1,50]$. For the binned setting, this yields in-distribution evaluation inside the two training bins, interpolation on $\xi \in (15,35)$, and extrapolation on $\xi \in [1,5) \cup (45,50]$.

\begin{figure}
    \centering
    \includegraphics[width=0.9\linewidth]{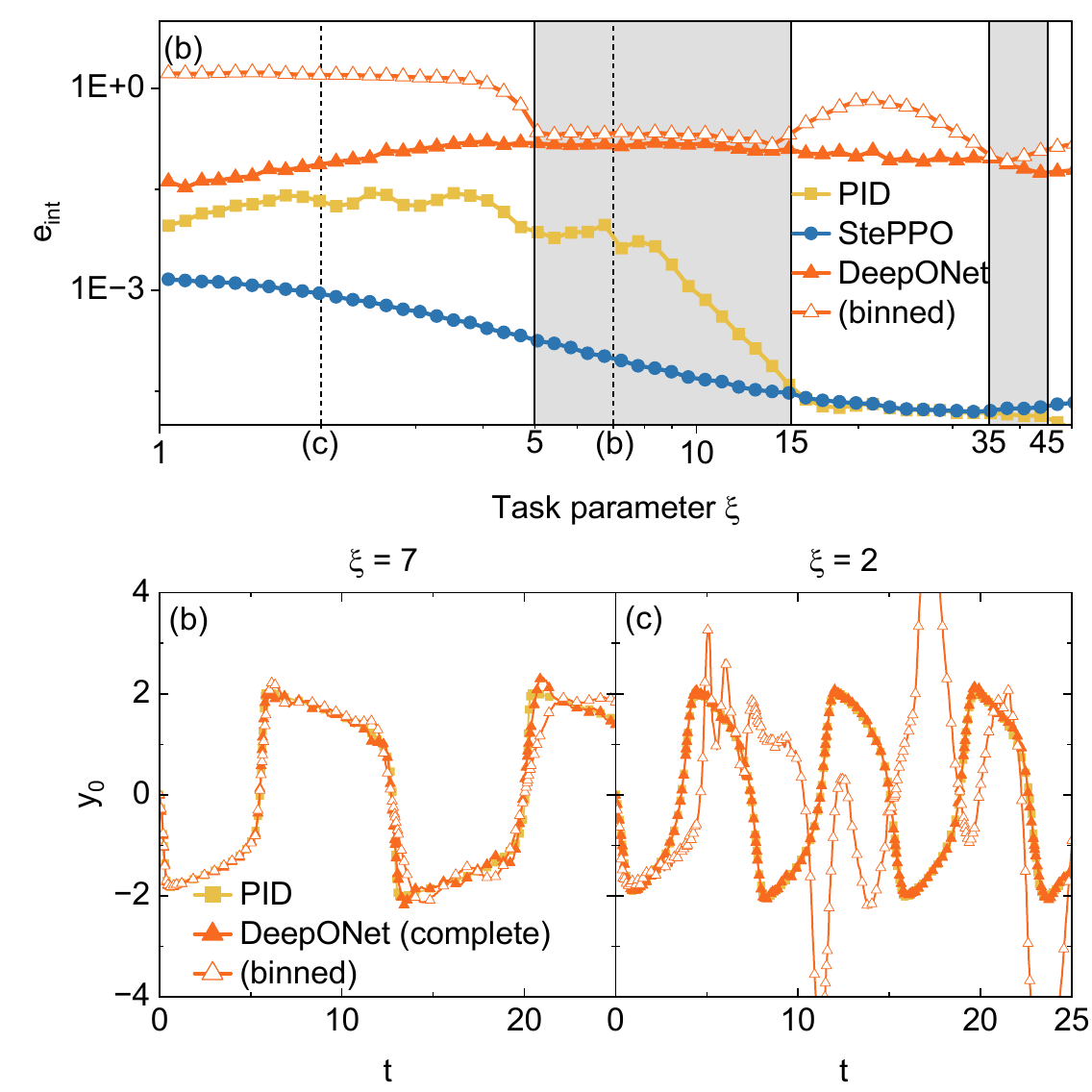}
    \caption{Robustness on the Van der Pol system under two training protocols. A \emph{full-range} model is trained on $\xi \in [1,50]$, whereas a \emph{binned} model is trained only on $\xi \in [5,15] \cup [35,45]$. All models are evaluated on $\xi \in [1,50]$. In panel (a), the shaded regions indicate the binned training intervals. We report the integrated error for PID, StePPO, and DeepONet; DeepONet is shown for both the full-range and binned training protocols. StePPO was also trained under both protocols, but the resulting curves were nearly identical, so only one is shown. Panels (b) and (c) show example trajectories for $\xi=7$ (inside the binned training support) and $\xi=2$ (outside the binned training support), respectively, comparing PID with DeepONet trained on the full range and on the binned support. For visualisation purposes, we set $t_{\text{end}}=25$ and show only every $5^{\text{th}}$ data point.}
    \label{fig:steps_vs_mu_vdp_ood}
\end{figure}

Figure~\ref{fig:steps_vs_mu_vdp_ood} reports the integrated error across the full evaluation range. For DeepONet, we show both the full-range and binned variants. StePPO was also trained under both protocols; however, the resulting curves were nearly identical, so only one StePPO curve is shown for readability. StePPO achieves accuracy comparable to PID across the full evaluation range, whereas DeepONet degrades markedly when training is restricted to the binned support, particularly outside the training bins.

\subsection{Study on the wall-clock time.}
Fewer numerical steps do not necessarily reduce runtime because learned controllers add a neural-network evaluation at each solver step. We therefore measure end-to-end wall-clock time for PID, StePPO, and DeepONet (Fig.~\ref{fig:auc}), excluding JIT compilation and using warm-up runs. PID and StePPO runtimes track their solver-step counts. Although StePPO evaluates a policy every iteration, it is substantially faster than PID over most task parameters, showing that step-count reductions outweigh inference overhead. For scalability, we repeat the experiment with $n_{\mathrm{env}} = 2^i$, $i \in \{1,2,3,4,5,6\}$. Batch latency stays roughly constant, so throughput increases with $n_{\mathrm{env}}$ via batched GPU execution, at the cost of higher GPU memory use. DeepONet behaves differently: it predicts the solution with a fixed number of network evaluations, so its inference time is largely independent of adaptive step count and nearly constant across task parameters. This highlights the inference-time efficiency of a fully learned surrogate, separate from accuracy and generalisation.

\input{figures/5_auc}

\section{Discussion}
The results suggest that learned control is most beneficial for trajectories with state-dependent stiff regions, where trajectory information can help anticipate step-size changes before repeated rejections occur. This is consistent with the larger PID--StePPO separation for Van der Pol and Brusselator and the smaller gain for scalar decay.

In contrast to fully data-driven approaches—such as surrogate models like DeepONet—two key differences stand out. To begin with, using a surrogate model can enable inference that is orders of magnitude faster. However, these models often have difficulty capturing trajectories accurately, particularly in systems with sharp transients or with widely varying task parameters. In addition, their performance typically degrades when asked to predict out-of-distribution data. Neither increasing the model size nor training for longer led to meaningful improvements. Since these models demand substantial computational resources for training while inference is often carried out on comparatively narrow or simple data, their practical applicability remains limited.


We have found that training dynamics are sensitive to the distribution of \say{task difficulty}. We note that different ranges in the binned training produced different convergences for Brusselator. Given $A=1$ and $B \in [1, 50]$, there is a large difference in trajectory dynamics. Without a proper choice of sampling-bin ranges and distribution, models could fail to learn properly, even in-distribution. We initially sampled $\log B \sim \mathcal{U}(\log 1,\log 100)$. However, given $A=1$, the hardest tasks were around $B=20$, and this distribution tended to oversample the easy tasks, causing a complete failure of training. As such, for future research we are interested in applying a sampling strategy that better targets harder tasks from a given task distribution \cite{greenberg2023trainhard}.

\section{Conclusion}
We formulated adaptive step-size control for parametrised initial value problems as a model-free Markov decision process and trained a continuous-action controller using PPO that preserves an embedded solver's local-error scaling and accept/reject rule. The policy augments classical error feedback with solution state and acceptance statistics. Results show little improvement on smooth scalar decay, but substantial reductions in solver attempts and wall-clock time for Van der Pol and Brusselator at matched tolerance, including held-out task parameters. This positions learned control as an extension of classical adaptive integration when avoided solver evaluations outweigh policy-inference cost.

\bibliography{references}
\bibliographystyle{IEEEtran}

\end{document}

%% file: figures/1_figure_method.tex
\begin{figure*}[t]
    \centering

    \begin{minipage}[c]{0.49\textwidth}
        \centering
        \input{figures/1_step_size_plot}\\[0.6ex]
        {\footnotesize (a) $\dot{y}(t)=-\lambda\cdot y(t)$ with $\lambda=14$}
    \end{minipage}
    \hfill
    \begin{minipage}[c]{0.49\textwidth}
        \centering
        \includegraphics[
            width=\linewidth,
            height=0.52\textwidth,
            keepaspectratio
        ]{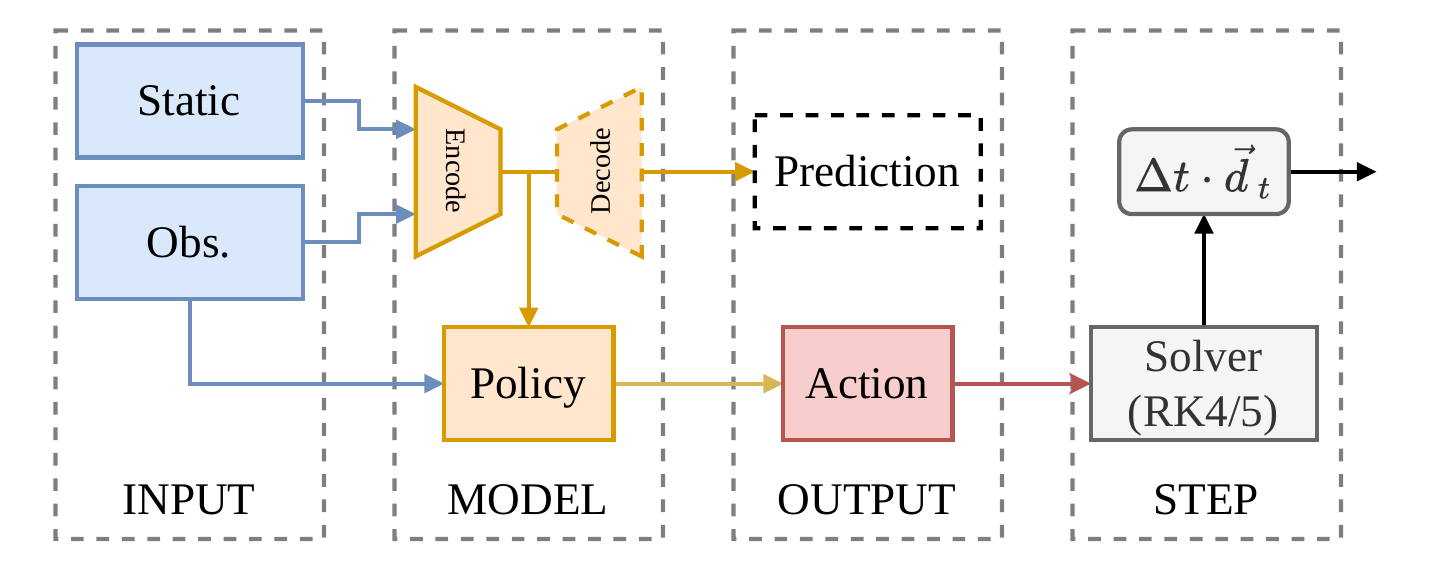}\\[0.5ex]
        {\footnotesize (b) Schematic of proposed model}
    \end{minipage}

    \caption{%
        Overview of the adaptive time-stepping problem and proposed controller. (a) Integral solve of $\dot{y}(t)=-\lambda\cdot y(t)$. We can write this expression analytically as $y(t) = y_0 e^{-\lambda t}$, providing a ground-truth trajectory. Red chords denote the steps taken by an elementary controller, whereas orange dashed chords indicate optimal step $h^\star$, the largest step from the same state whose embedded error estimate still satisfies the prescribed tolerance. This indicates the potential of an improved controller.
        (b) Basic principle of the learned adaptive time-step controller. State and static information are supplied to the encoder and policy to determine the subsequent step-size action, which is used by the embedded solver.}
    \label{fig:method}
\end{figure*}

%% file: figures/1_step_size_plot.tex
\definecolor{stepInk}{HTML}{1A1A1A}
\definecolor{stepExact}{HTML}{4D4D4D}
\definecolor{stepState}{HTML}{1F6FB2}
\definecolor{stepTaken}{HTML}{C0392B}
\definecolor{stepMax}{HTML}{D98218}
\definecolor{stepTol}{HTML}{C9A227}
\definecolor{stepTangent}{HTML}{8A9198}

\pgfplotsset{
  stepaxis/.style={
    axis line style={stepInk, line width=0.3pt},
    tick style={stepInk, line width=0.3pt},
    tick label style={font=\footnotesize},
    label style={font=\footnotesize},
    every axis plot/.append style={line join=round},
    scale only axis,
  },
  exactline/.style={stepExact, line width=0.5pt, densely dotted},
  takenline/.style={stepTaken, line width=1.0pt},
  maxline/.style={stepMax, line width=1.0pt, dashed},
  tangentline/.style={stepTangent, line width=0.5pt, densely dashed},
  dirline/.style={stepTangent, line width=1.0pt},
  dtbar/.style={stepInk, line width=0.7pt},
  statemark/.style={stepState, only marks, mark=*, mark size=1.4pt,
                    mark options={fill=stepState, draw=white, line width=0.3pt}},
  lowermark/.style={stepState, only marks, mark=o, mark size=1.4pt,
                    mark options={draw=stepState, line width=0.6pt}},
  maxmark/.style={stepMax, only marks, mark=*, mark size=2.2pt,
                  mark options={fill=stepMax, draw=white, line width=0.3pt}},
  maxlowermark/.style={stepMax, only marks, mark=o, mark size=1.4pt,
                       mark options={draw=stepMax, line width=0.6pt}},
  tolbars/.style={stepTol, only marks, mark=none,
                  error bars/y dir=both,
                  error bars/y explicit,
                  error bars/error bar style={stepTol, line width=0.8pt},
                  error bars/error mark=-,
                  error bars/error mark options={stepTol, line width=0.8pt,
                                                 mark size=1.6pt}},
}

\begin{tikzpicture}
\begin{axis}[
    stepaxis, name=mainaxis,
    width=66mm, height=36mm,
    legend columns=1,
    legend cell align=left,
    legend style={at={(0.99,0.98)}, anchor=north east, draw=none,
                  font=\scriptsize},
    xlabel={$t$ (a.u.)}, ylabel={$y$ (a.u.)},
    xmin=-0.02, xmax=1.02, ymin=-0.1, ymax=1.12,
    xtick={0,0.25,0.5,0.75,1}, ytick={0,0.5,1},
  ]
  \addplot[exactline, domain=0:1.02, samples=201] {exp(-14*x)};
  \addlegendentry{$y(t)$}

  \addplot[takenline] coordinates {
    (0.000000,1.000000) (0.000100,0.998601) (0.000200,0.997204)
    (0.001200,0.983340) (0.011200,0.854875) (0.111200,0.211284)
    (0.204386,0.057403) (0.252207,0.029390) (0.304546,0.014125)
  };
  \addlegendentry{$h_n$}

  \addplot[statemark] coordinates {
    (0.000000,1.000000) (0.000100,0.998601) (0.000200,0.997204)
    (0.001200,0.983340) (0.011200,0.854875) (0.111200,0.211284)
    (0.204386,0.057403) (0.252207,0.029390) (0.304546,0.014125)
  };
  \addlegendentry{$y_n$}

  \addplot[maxline] coordinates {
    (0.252207,0.029390) (0.372986,0.005454)
  };
  \addlegendentry{$h^\star$}
  \addplot[maxmark, forget plot] coordinates {
    (0.372986,0.005454)
  };

  \draw[stepTangent, densely dotted, line width=0.4pt]
    (axis cs:0.304546,-0.1) -- (axis cs:0.304546,0.014125);
  \draw[stepMax, densely dotted, line width=0.4pt]
    (axis cs:0.372986,-0.1) -- (axis cs:0.372986,0.005454);

  \addlegendimage{dtbar}
  \addlegendentry{$\Delta t_n$}

  \coordinate (zNW) at (axis cs:0.235,0.05);
  \coordinate (zNE) at (axis cs:0.40,0.05);
  \coordinate (zSE) at (axis cs:0.40,-0.035);
\end{axis}
\draw[stepInk, line width=0.28pt, densely dashed] (zNW) rectangle (zSE);

\begin{axis}[
    stepaxis, name=insetaxis,
    at={(mainaxis.north west)}, anchor=north west, xshift=12mm, yshift=-4mm,
    width=30mm, height=16mm,
    xmin=0.235, xmax=0.40, ymin=-0.035, ymax=0.05,
    tick label style={font=\tiny}, clip=true,
    axis background/.style={fill=white},
  ]
  \addplot[exactline, domain=0.237713:0.403181, samples=100]
    {exp(-14*x)};
  \addplot[maxline]  coordinates {
    (0.252207,0.029390) (0.372986,0.005454)
  };
  \addplot[takenline] coordinates {
    (0.252207,0.029390) (0.304546,0.014125)
  };
    

    \draw[stepInk, line width=0.7pt]
        (axis cs:0.252207,0) --
        node[midway, fill=white, inner sep=0.7pt, font=\scriptsize]
            {$\Delta t_n$}
        (axis cs:0.304546,0);
    
    \draw[stepInk, line width=0.7pt]
        (axis cs:0.252207,-0.002) -- (axis cs:0.252207,0.002)
        (axis cs:0.304546,-0.002) -- (axis cs:0.304546,0.002);

    \draw[stepInk, line width=0.7pt]
        (axis cs:0.252207,-0.01) --
        node[midway, fill=white, inner sep=0.7pt, font=\scriptsize]
            {$\Delta t^\star$}
        (axis cs:0.372986,-0.01);
    
    \draw[stepInk, line width=0.7pt]
        (axis cs:0.252207,-0.012) -- (axis cs:0.252207,-0.008)
        (axis cs:0.372986,-0.012) -- (axis cs:0.372986,-0.008);

  \addplot[lowermark]    coordinates {
    (0.304546,0.014124)
  };
  \addplot[maxlowermark] coordinates {
    (0.372986,0.005424)
  };
  \addplot[maxmark]      coordinates {
    (0.372986,0.005454)
  };
  \addplot[statemark] coordinates {
    (0.252207,0.029390) (0.304546,0.014125)
  };
\end{axis}

\draw[stepInk, line width=0.2pt, densely dotted] (zNW) -- (insetaxis.south west);
\draw[stepInk, line width=0.2pt, densely dotted] (zNE) -- (insetaxis.south east);
\end{tikzpicture}

%% file: figures/2_trajectories.tex
%

\definecolor{pidcolor}{RGB}{150,150,40}      
\definecolor{steppocolor}{RGB}{90,160,220}   
\definecolor{deepocolor}{RGB}{230,140,20}    

\pgfplotsset{
  pidstyle/.style={pidcolor, thick, mark=o, mark size=1.6pt, mark repeat=25},
  steppostyle/.style={steppocolor, thick, mark=o, mark size=1.6pt, mark repeat=25},
  deepostyle/.style={deepocolor, thick, mark=o, mark size=1.6pt, mark repeat=25},
}

\begin{figure}[t]
\centering

\begin{tikzpicture}
\begin{groupplot}[
    group style={
        group size=1 by 8,
        vertical sep=0pt,
        xlabels at=edge bottom,
    },
    width=0.90\columnwidth,
    height=35mm,
    tick label style={font=\scriptsize},
    label style={font=\scriptsize},
]

\nextgroupplot[
    xmin=0, xmax=10, ylabel={$y$}, 
    xticklabels=\empty, 
    legend to name=figlegend, legend columns=-1,
    legend style={font=\scriptsize},
]
\node[anchor=north west] at (rel axis cs:0.02,0.95) {\scriptsize (a)};
\addplot[pidstyle]    table[col sep=comma, x=t, y=PID_y] {data/scalar_decay_pid_rl_deeponet_trajectory.csv};
\addlegendentry{PID}
\addplot[steppostyle]  table[col sep=comma, x=t, y=RL_y]  {data/scalar_decay_pid_rl_deeponet_trajectory.csv};
\addlegendentry{StePPO}
\addplot[deepostyle]  table[col sep=comma, x=t, y=D_y]   {data/scalar_decay_pid_rl_deeponet_trajectory.csv};
\addlegendentry{DeepONet}

\nextgroupplot[
    xmin=0, xmax=10, ylabel={$e$}, xlabel={$t$},
    ymin=-4, ymax=2, ytick={-4,-2,0,2},
    yticklabel={$10^{\pgfmathprintnumber[precision=0]{\tick}}$},
]
\node[anchor=south west] at (rel axis cs:0.02,0.95) {\scriptsize (b)};
\addplot[black, dashed, no marks] coordinates {(0,0) (10,0)};
\addplot[pidstyle]   table[col sep=comma, x=t, y=PID_err] {data/scalar_decay_pid_rl_deeponet_trajectory.csv};
\addplot[steppostyle] table[col sep=comma, x=t, y=RL_err]  {data/scalar_decay_pid_rl_deeponet_trajectory.csv};
\addplot[deepostyle]  table[col sep=comma, x=t, y=D_err]   {data/scalar_decay_pid_rl_deeponet_trajectory.csv};

\nextgroupplot[
    xmin=0, xmax=50, ylabel={$y_1$}, xticklabels=\empty, yshift=-0.6cm
]
\node[anchor=north west] at (rel axis cs:0.02,0.95) {\scriptsize (a)};
\addplot[pidstyle]    table[col sep=comma, x=t, y=PIDu_y0] {data/van_der_pol_pid_rl_deeponet_trajectory.csv};
\addplot[steppostyle]  table[col sep=comma, x=t, y=RL_y0]   {data/van_der_pol_pid_rl_deeponet_trajectory.csv};
\addplot[deepostyle]  table[col sep=comma, x=t, y=D_y0]    {data/van_der_pol_pid_rl_deeponet_trajectory.csv};

\nextgroupplot[xmin=0, xmax=50, ylabel={$y_2$}, xticklabels=\empty]
\node[anchor=north west] at (rel axis cs:0.02,0.95) {\scriptsize (b)};
\addplot[pidstyle]    table[col sep=comma, x=t, y=PIDu_y1] {data/van_der_pol_pid_rl_deeponet_trajectory.csv};
\addplot[steppostyle]  table[col sep=comma, x=t, y=RL_y1]   {data/van_der_pol_pid_rl_deeponet_trajectory.csv};
\addplot[deepostyle]  table[col sep=comma, x=t, y=D_y1]    {data/van_der_pol_pid_rl_deeponet_trajectory.csv};

\nextgroupplot[
xmin=0, xmax=50, ylabel={$e$}, xlabel={$t$},
ymin=-4, ymax=2, ytick={-4,-2,0,2},
yticklabel={$10^{\pgfmathprintnumber[precision=0]{\tick}}$},
]
\node[anchor=south west] at (rel axis cs:0.02,0.95) {\scriptsize (c)};
\addplot[black, dashed, no marks] coordinates {(0,0) (50,0)};
\addplot[pidstyle]    table[col sep=comma, x=t, y=PIDu_err] {data/van_der_pol_pid_rl_deeponet_trajectory.csv};
\addplot[steppostyle]  table[col sep=comma, x=t, y=RL_err]   {data/van_der_pol_pid_rl_deeponet_trajectory.csv};
\addplot[deepostyle]  table[col sep=comma, x=t, y=D_err]    {data/van_der_pol_pid_rl_deeponet_trajectory.csv};

\nextgroupplot[
    xmin=0, xmax=100, ylabel={$y_1$}, xticklabels=\empty, yshift=-0.6cm
]
\node[anchor=north west] at (rel axis cs:0.02,0.95) {\scriptsize (a)};
\addplot[pidstyle]    table[col sep=comma, x=t, y=PIDu_y0] {data/brusselator_pid_rl_deeponet_trajectory.csv};
\addplot[steppostyle]  table[col sep=comma, x=t, y=RL_y0]   {data/brusselator_pid_rl_deeponet_trajectory.csv};
\addplot[deepostyle]  table[col sep=comma, x=t, y=D_y0]    {data/brusselator_pid_rl_deeponet_trajectory.csv};

\nextgroupplot[
    xmin=0, xmax=100, ylabel={$y_2$}, xticklabels=\empty
]
\node[anchor=north west] at (rel axis cs:0.02,0.95) {\scriptsize (b)};
\addplot[pidstyle]    table[col sep=comma, x=t, y=PIDu_y1] {data/brusselator_pid_rl_deeponet_trajectory.csv};
\addplot[steppostyle]  table[col sep=comma, x=t, y=RL_y1]   {data/brusselator_pid_rl_deeponet_trajectory.csv};
\addplot[deepostyle]  table[col sep=comma, x=t, y=D_y1]    {data/brusselator_pid_rl_deeponet_trajectory.csv};

\nextgroupplot[
    xmin=0, xmax=100, ylabel={$e$}, xlabel={$t$},
    ymin=-4, ymax=2, ytick={-4,-2,0,2},
    yticklabel={$10^{\pgfmathprintnumber[precision=0]{\tick}}$},
]
\node[anchor=south west] at (rel axis cs:0.02,0.95) {\scriptsize (c)};
\addplot[black, dashed, no marks] coordinates {(0,0) (100,0)};
\addplot[pidstyle]    table[col sep=comma, x=t, y=PIDu_err] {data/brusselator_pid_rl_deeponet_trajectory.csv};
\addplot[steppostyle]  table[col sep=comma, x=t, y=RL_err]   {data/brusselator_pid_rl_deeponet_trajectory.csv};
\addplot[deepostyle]  table[col sep=comma, x=t, y=D_err]    {data/brusselator_pid_rl_deeponet_trajectory.csv};

\end{groupplot}

\node at ($(group c1r1.north)+(0,-0.55cm)$) {\ref{figlegend}};

\end{tikzpicture}

\caption{The state trajectory and the corresponding local error estimate given by (\ref{eq:e_i}). The max error threshold is set at $e_{\text{max}}=1$, where $e_{i} > 1$ is rejected. Note how for the scalar decay, the non-stiff problems (small $q$) operate far below the error threshold, while highly damped system (initially) operate near error threshold. Also note that DeepONet operates orders of magnitude above the error threshold than acceptable for adaptive time steppers.
}
\label{fig:trajectories}

\end{figure}

%% file: figures/3_figure.tex
\begin{figure*}[t]
    \centering
    \begin{minipage}{\textwidth}
        \centering
        \includegraphics[
            width=\textwidth,
            trim={0 0 0 0},
            clip
        ]{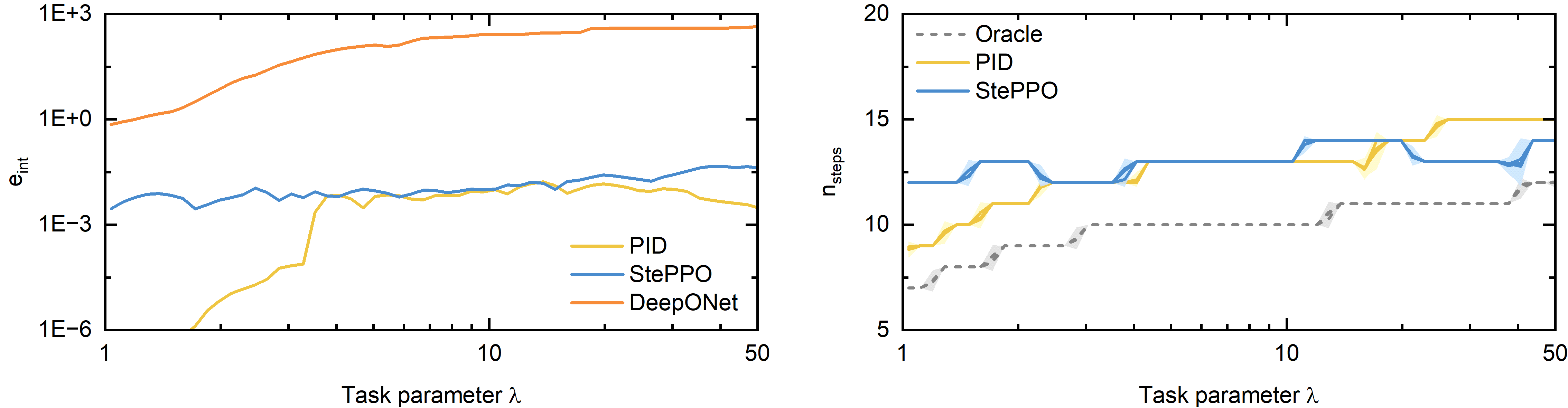}\\[-0.3ex]
        {\footnotesize (a) Scalar decay}
    \end{minipage}

    \vspace{0.8ex}

    \begin{minipage}{\textwidth}
        \centering
        \includegraphics[
            width=\textwidth,
            trim={0 0 0 0},
            clip
        ]{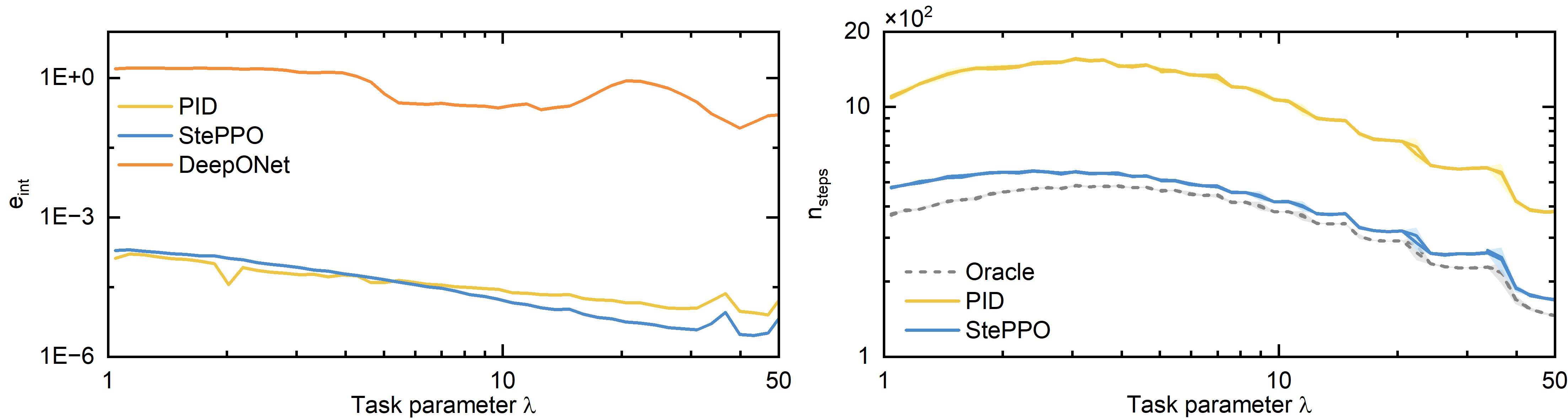}\\[-0.3ex]
        {\footnotesize (b) Van der Pol}
    \end{minipage}

    \vspace{0.8ex}

    \begin{minipage}{\textwidth}
        \centering
        \includegraphics[
            width=\textwidth,
            trim={0 0 0 0},
            clip
        ]{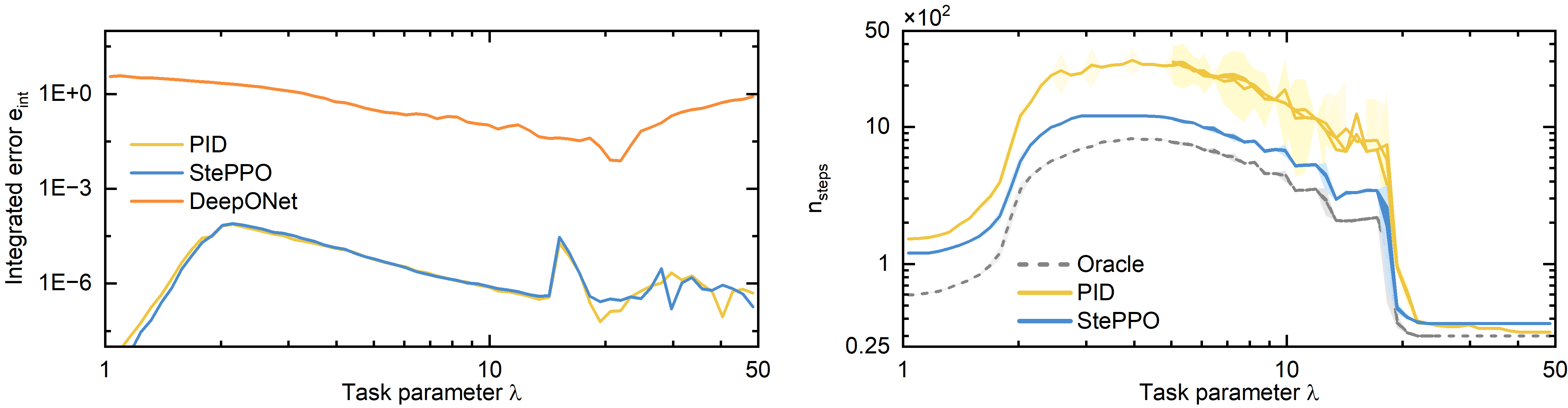}\\[-0.3ex]
        {\footnotesize (c) Brusselator}
    \end{minipage}

    \caption{%
        Performance across the three benchmark systems. Left column: integrated error $e_{\mathrm{int}}$ with respect to the reference trajectory. Right column: number of solver steps $n_{\mathrm{steps}}$ required by PID and StePPO. The Oracle indicates a lower bound on the number of steps required for the corresponding task parameter $\lambda$. Rows show (a) scalar decay, (b) Van der Pol, and (c) Brusselator.
    }
    \label{fig:results}
\end{figure*}

%% file: figures/5_auc.tex
\begin{figure}
\centering
\begin{tikzpicture}
\definecolor{cPID}{HTML}{F86A1F}
\definecolor{cStePPO}{HTML}{6FA9DA}
\definecolor{cDeepONet}{HTML}{E8C047}

\begin{axis}[
    width=85mm, height=7cm,
    xmode=log,
    xlabel={Task parameter $\xi$},
    ylabel={Wall-clock (ms)},
    xmin=1,xmax=200,
    legend pos=north east,
    grid=major,
]

\addplot[cPID, mark=*, mark size=1.5pt, thick, name path=A]
    table[col sep=comma, x=x, y=PID] {data/vdp_vae_vdp_3_batch1_gpu_wallclock_vs_mu_combined.csv};
\addlegendentry{PID}

\addplot[cStePPO, mark=*, mark size=1.5pt, thick, name path=B]
    table[col sep=comma, x=x, y=RL] {data/vdp_vae_vdp_3_batch1_gpu_wallclock_vs_mu_combined.csv};
\addlegendentry{StePPO}

\addplot[cDeepONet, mark=*, mark size=1.5pt, thick, name path=C]
    table[col sep=comma, x=x, y=DeepONet] {data/vdp_vae_vdp_3_batch1_gpu_wallclock_vs_mu_combined.csv};
\addlegendentry{DeepONet}

\addplot[cPID!50!cStePPO, opacity=0.25] fill between[of=A and B];

\addplot[cStePPO!50!cDeepONet, opacity=0.25] fill between[of=B and C];

\end{axis}
\end{tikzpicture}

\caption{Wall-clock (ms) vs $\xi$ for Van der Pol for $n_{\text{env}}=1$ on GPU using PID, StePPO and DeepONet. Indicated shaded regions show the wall-clock reduction of RL over PID and the remaining reduction of DeepONet over RL. DeepONet maintains a nearly constant wall-clock time.}
\label{fig:auc}
\end{figure}
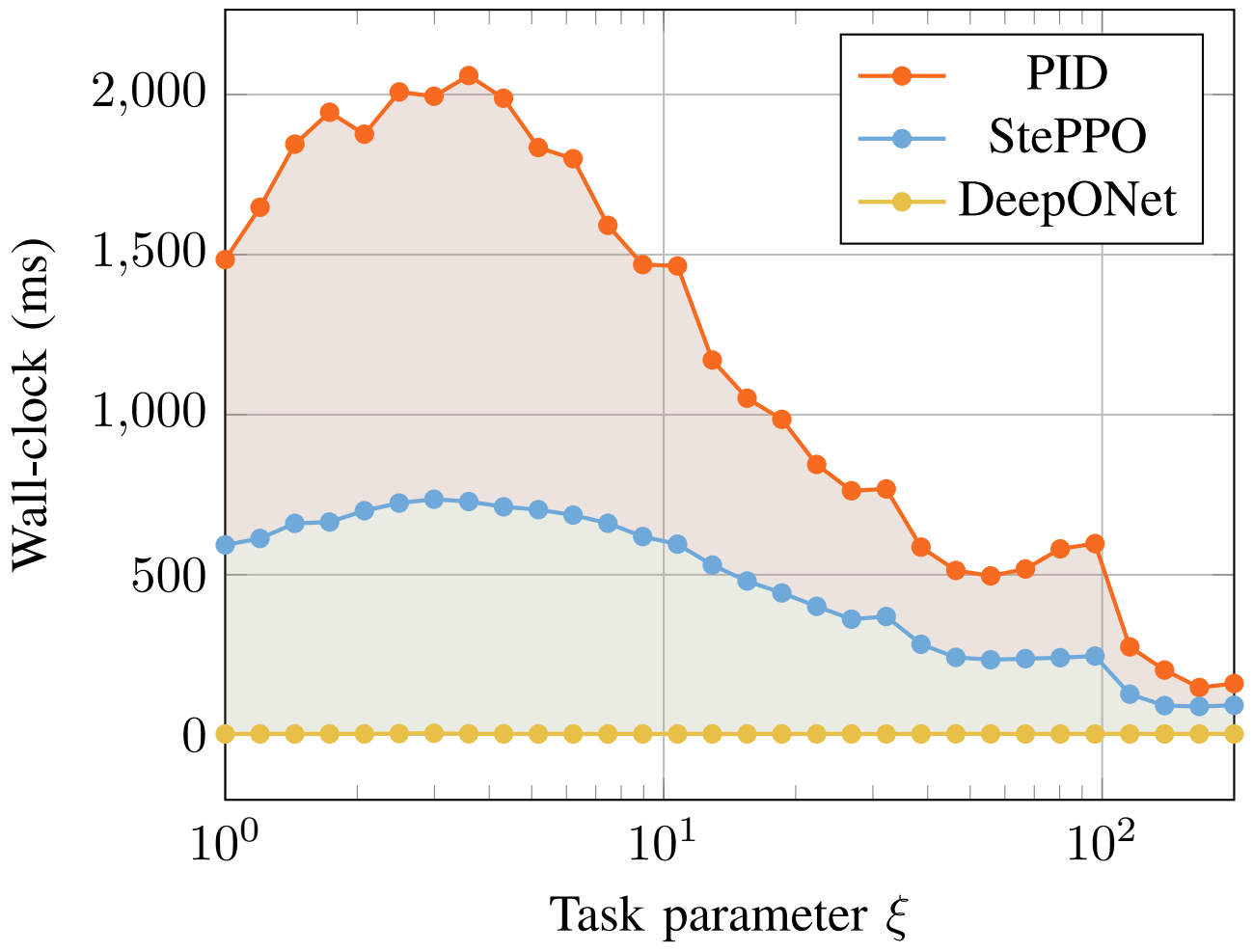